\pdfoutput=1
\documentclass{article}
\usepackage{spconf,amsmath,graphicx}

\usepackage[utf8]{inputenc}
\usepackage{float}
\usepackage[symbol,bottom,hang,flushmargin]{footmisc}
\usepackage[acronym,toc,shortcuts,nonumberlist]{glossaries}
\usepackage[tracking=true]{microtype}
\usepackage{multirow}
\usepackage{siunitx}
\usepackage[usenames,dvipsnames]{xcolor}
\usepackage{xspace}
\usepackage{adjustbox}
\usepackage[normalem]{ulem}
\usepackage{hyperref}
\usepackage[skip=1pt,margin=0pt]{caption}

\makeatletter

\renewcommand{\section}{\@startsection
  {section}%
  {1}%
  {}%
  {0.5\baselineskip}%
  {0.5\baselineskip}%
  {}}%

\renewcommand{\subsection}{\@startsection
  {subsection}%
  {2}%
  {}%
  {0.5\baselineskip}%
  {0.5\baselineskip}%
  {}}%

\renewcommand{\subsubsection}{\@startsection
  {subsubsection}%
  {3}%
  {}%
  {0.5\baselineskip}%
  {0.5\baselineskip}%
  {}}%

\g@addto@macro\normalsize{%
  \setlength\abovedisplayskip{4pt plus 2pt minus 2pt}
  \setlength\belowdisplayskip{4pt plus 2pt minus 2pt}
  \setlength\abovedisplayshortskip{4pt plus 2pt minus 2pt}
  \setlength\belowdisplayshortskip{4pt plus 2pt minus 2pt}
}

\makeatother

\newacronym{AM}{AM}{acoustic model}
\newacronym{ASR}{ASR}{automatic speech recognition}
\newacronym[longplural={bi-directional long-short term memories}]{BLSTM}{BLSTM}{bi-directional long-short term memory}
\newacronym{CART}{CART}{classification and regression tree}
\newacronym{CE}{CE}{cross entropy}
\newacronym{CNN}{CNN}{convolutional neural network}
\newacronym{DNN}{DNN}{deep neural network}
\newacronym{DNN-HMM}{DNN-HMM}{deep neural network hidden Markov model}
\newacronym{fCE}{fCE}{frame-wise cross-entropy}
\newacronym{GMM}{GMM}{Gaussian mixture model}
\newacronym{GMM-HMM}{GMM-HMM}{Gaussian mixture model hidden Markov model}
\newacronym{GPU}{GPU}{graphics processing unit}
\newacronym{HMM}{HMM}{hidden Markov model}
\newacronym{LM}{LM}{language model}
\newacronym[longplural={long-short term memories}]{LSTM}{LSTM}{long-short term memory}
\newacronym{NN}{NN}{neural network}
\newacronym{WER}{WER}{word error rate}
\newacronym{WERR}{WERR}{word error rate reduction}

\newcommand{\tab}{Table~}

\newcommand{\sect}{Section~}
\newcommand{\wvtwo}{\textit{wav2vec 2.0}\xspace}  %
\newcommand{\xlsr}{\textit{XLSR-53}\xspace}
\newcommand{\lrg}{\textit{Large}\xspace}  %
\newcommand{\lrgcut}{\textit{Large}\textsubscript{1-8}\xspace}
\newcommand{\lrgcutn}{\textit{Large}\textsubscript{1-N}\xspace}

\newcommand{\lrgcuteighteen}{\textit{Large}\textsubscript{1-18}\xspace}

\newcommand{\xlsrcut}{\textit{XLSR-53}\textsubscript{1-8}\xspace}
\newcommand{\hykist}{HYKIST\xspace}
\newcommand{\datahykist}[1][ ]{$\mathcal{D}_{\textrm{H}}^{\textrm{#1}}$\xspace}
\newcommand{\datainhouse}[1][ ]{$\mathcal{D}_{\textrm{ih}}^{\textrm{#1}}$\xspace}
\newcommand{\datainhouseml}{$\mathcal{D}_{\textrm{ih}}^{\textrm{VGA}}$\xspace}

\title{Efficient Utilization of Large Pre-Trained Models for Low Resource ASR}
\name{
Peter Vieting$^{*1}$, Christoph L\"uscher$^{*1,2}$, Julian Dierkes$^{1}$, Ralf Schl\"uter$^{1,2}$, Hermann Ney$^{1,2}$\thanks{$^{*}$equal contribution}
}
\address{
$^1$Human Language Technology and Pattern Recognition Group,\\
Computer Science Department, RWTH Aachen University, 52074 Aachen, Germany\\
$^2$AppTek GmbH, 52062 Aachen, Germany\\
\textit{\{vieting,luescher,schlueter,ney\}@cs.rwth-aachen.de}
}
\begin{document}
\maketitle
\begin{abstract}
Unsupervised representation learning has recently helped \gls{ASR} to tackle tasks with limited labeled data.
Following this, hardware limitations and applications give rise to the question how to take advantage of large pre-trained models efficiently and reduce their complexity.
In this work, we study a challenging low resource conversational telephony speech corpus from the medical domain in Vietnamese and German.
We show the benefits of using unsupervised techniques beyond simple fine-tuning of large pre-trained models, discuss how to adapt them to a practical telephony task including bandwidth transfer and investigate different data conditions for pre-training and fine-tuning.
We outperform the project baselines by 22\% relative using pre-training techniques.
Further gains of 29\% can be achieved by refinements of architecture and training and 6\% by adding \SI{0.8}{\hour} of in-domain adaptation data.

\end{abstract}
\section{Introduction}
\label{sec:intro}

The development of \gls{ASR} systems has come a long way and established remarkable performance, especially on tasks with sufficient training data.
However, varying acoustic and recording conditions and speaking styles as well as a lack of sufficient in-domain training data still pose challenges to the development of accurate models \cite{cui2015babel}.
Unsupervised pre-training has recently allowed to exploit unlabeled audio data which is available at much lower cost, significantly reducing the need for transcribed data.
Additionally, the public availability of pre-trained model checkpoints is appealing to reduce training resource consumption both from an economical and environmental point of view.

Nevertheless, these models are often very large, requiring cutting-edge hardware both for training and recognition to satisfy the computational and memory requirements.
Moreover, application requirements regarding the real time factor in recognition can be difficult to meet.
This gives rise to the question, how to efficiently take advantage of large pre-trained models and how to reduce their complexity in order to meet the demands mentioned above.

Furthermore, despite the feasibility of training \gls{ASR} systems on very small amounts of labeled data when using pre-trained models, there is certainly room for improvement beyond vanilla fine-tuning of existing models.
This paper addresses a challenging real-world low-resource task.
Concretely, we use a conversational telephony speech corpus from the medical domain with very small amount of data in Vietnamese and German.
This task constitutes a prime example for the application of pre-trained models while still posing several challenges like domain shift regarding the unsupervised models' training data (conversational speech, acoustic conditions, medical domain), telephony bandwidth and application requirements on limiting the complexity of models and training.

This work shows how to exploit large pre-trained models in a practical scenario with limited resources and has contributions along three main lines.
The \textbf{sampling rate mismatch} is addressed beyond simple re-sampling by different proposed modifications of the feature extractor.
We \textbf{reduce model sizes} and GPU memory footprint by exploiting intermediate representations and applying freezing schemes.
Moreover, we study multi-stage pre-training and fine-tuning to address the \textbf{data conditions and achieve adaptation} for the target task.

\section{Related work}
\label{sec:related}

Unsupervised approaches have gained popularity since they have shown a potential of high performance with only little annotated data \cite{mohamed2022representation_review}.
Initial work applied this method to an \gls{ASR} task by running unsupervised pre-training on a large unlabeled dataset, followed by a fine-tuning step with a small annotated dataset \cite{deepmind2018cpc, facebook2019wav2vec, facebook2020wav2vec2}.
This technique can drastically reduce the amount of labeled data which is necessary to build \gls{ASR} systems.
The successes motivated further research into improving the modeling approach \cite{facebook2021hubert, facebook2022wav2vecaug} and understanding the individual components \cite{livescu2021wav2vec_analysis}.
Furthermore, the data used for pre-training and fine-tuning was studied, e.g., in a domain-shift scenario \cite{robust_wav2vec2} or using multilingual data \cite{facebook2020xlsr}.

Since the unsupervised loss is computed solely based on the input speech audio without any need for labels, it is particularly appealing in a multilingual scenario.
It is straight-forward to apply it for different languages or entirely multilingual data.
A number of papers have started investigating this research direction \cite{facebook2020xlsr, microsoft2021unispeech, zhang2021xlst, google2022just}.
Before, supervised training with multilingual data could also show improvements for low resource languages by using a separate output layer per language providing speech representations transferable to languages unseen in multilingual training \cite{tuske2014multilingual}.

Reducing the complexity of large pre-trained models was studied in the literature, e.g. using knowledge distillation \cite{2021_shrinking_bigfoot, taiwan2022distilhubert}.
Other works suggest that the information learned in intermediate layers is more related to what is helpful for \gls{ASR} and related tasks \cite{livescu2021wav2vec_analysis}, which motivates using these layers as output instead of the whole model allowing to reduce the model size at the same time.

The corpus used in this paper originates from the \hykist project\footnote{\href{https://www.bundesgesundheitsministerium.de/ministerium/ressortforschung-1/handlungsfelder/forschungsschwerpunkte/digitale-innovation/modul-4-smarte-kommunikation/hykist.html}{https://bit.ly/hykist\_project}}.
The data and work on baselines is presented in \cite{rwth2022hykist_baseline}, which also elaborates on the challenges of the medical domain.
We further extend the work by using unsupervised methods here, especially focusing on the question how to make use of large pre-trained models efficiently.

\section{Methods}
We follow the setup described in \cite{rwth2022hykist_baseline}.
It deploys a hybrid \gls{NN}-\gls{HMM} model based on a \gls{GMM}-\gls{HMM} alignment of speech and labels.
The lexica and 4gram \glspl{LM} used for all experiments are described in \cite{rwth2022hykist_baseline}.

For unsupervised training, a two-stage training setup is applied.
First, the \wvtwo framework is used to pre-train an \gls{NN} on unlabeled (monolingual or multilingual) data using the contrastive loss and diversity loss as described in \cite{facebook2020wav2vec2}.
Subsequently, a fine-tuning is conducted on target language data by initializing the \gls{NN} used in the \gls{AM} of the hybrid model with a checkpoint from pre-training, adding a softmax output layer and training with the \gls{fCE} loss using the alignment mentioned above as e.g. in \cite{vieting2021waveform}.
In addition to pre-training own models on our custom data, we also investigate exploiting a publicly available model, i.e., \xlsr \cite{facebook2020xlsr}.
This was pre-trained on \SI{56}{\kilo\hour} of speech data from 53 different languages for 800k steps (19 epochs) and we use the checkpoint that was not fine-tuned to any language\footnote{\href{https://github.com/facebookresearch/fairseq/tree/main/examples/wav2vec}{https://github.com/facebookresearch/fairseq/tree/main/examples/wav2vec}}.
Furthermore, we experiment with using the \xlsr model as an initialization for \wvtwo pre-training on our custom data followed by regular fine-tuning.

The 24 Transformer blocks used in the \lrg architecture impose high demands on the \gls{GPU} memory.
Trading \gls{GPU} memory against a decreased batch size leads to significantly longer training times.
Prior work has shown that representations from intermediate layers of pre-trained models contain information that is useful for \gls{ASR}, even more than the output of the final layers \cite{livescu2021wav2vec_analysis}.
Motivated by this observation, we propose to reduce the model size by cutting off the \wvtwo \lrg encoder after the $N^{th}$ Transformer block and refer to the model as \lrgcutn.
This can be done for pre-training already or for fine-tuning only.

\section{Experiments}
\label{sec:experiments}
In this section, we describe our experimental setups.
We use RETURNN%
~\cite{doetsch2016returnn}
for supervised training and Fairseq%
~\cite{facebook2019fairseq}
for unsupervised \wvtwo training.
Decoding is performed with RASR%
~\cite{rybach2011rasr}.
We convert the Fairseq models to RETURNN models with an automatic
conversion toolkit\footnote{\href{https://github.com/rwth-i6/pytorch-to-returnn-converter}{https://github.com/rwth-i6/pytorch-to-returnn-converter}}.
We plan to publish training and decoding configurations
online\footnote{\href{https://github.com/rwth-i6/i6-experiments}{https://github.com/rwth-i6/i6-experiments}}.

\subsection{Data}
\label{sec:data}
Within the \hykist project, a corpus of telephone conversations between patients, doctors and interpreters was recorded \cite{rwth2022hykist_baseline}.
We denote it as \datahykist here.
It is split into adapt, dev and test sets where the adaptation set contains \SI{0.8}{\hour} and \SI{4.7}{\hour} for Vietnamese and German, respectively.
Additionally, we use an annotated in-house dataset of \SI{8}{\kilo\hertz} sampled conversational telephone speech in each language (\SI{219}{\hour}/\SI{177}{\hour}) for fine-tuning and the combination of the audio data together with more Arabic data (\SI{768}{\hour}), which is the third considered language in the project.
We denote it as \datainhouse and represent the respective language by superscript V, G, A or VGA for the joint multilingual data.
Data statistics are presented in \cite{rwth2022hykist_baseline}.

\subsection{Investigation of Data Conditions in Pre-Training}

\begin{table*}[!htb]
\setlength{\tabcolsep}{0.3em}
\centering
\caption{
WERs {[}\%{]} for both Vietnamese and German.
Pre-trainings have been done with random initialization or using the public \xlsr checkpoint as initialization.
Pre-training "none" with random initialization means fine-tuning from scratch.
All fine-tunings use the \lrgcut architecture and are trained until full convergence on the in-house data of the respective language.
}
\begin{tabular}{|cccc|c|ccc|ccc|}
\hline
\multicolumn{4}{|c|}{Pre-training}                                                                                                                                                                                                                                 & Fine-tuning             & \multicolumn{3}{c|}{Viet. WER {[}\%{]}}                                         & \multicolumn{3}{c|}{Ger. WER {[}\%{]}}                                          \\ \hline
\multicolumn{1}{|c|}{\multirow{2}{*}{Architecture}} & \multicolumn{1}{c|}{\multirow{2}{*}{Init}}    & \multicolumn{1}{c|}{\multirow{2}{*}{Data (in-house)}}                                                                                 & \multirow{2}{*}{Epochs} & \multirow{2}{*}{Epochs} & \multicolumn{2}{c|}{\hykist}                           & \multirow{2}{*}{In-house} & \multicolumn{2}{c|}{\hykist}                           & \multirow{2}{*}{In-house} \\ \cline{6-7} \cline{9-10}
\multicolumn{1}{|c|}{}                              & \multicolumn{1}{c|}{}                         & \multicolumn{1}{c|}{}                                                                                                              &                         &                         & \multicolumn{1}{c|}{dev}  & \multicolumn{1}{c|}{test} &                         & \multicolumn{1}{c|}{dev}  & \multicolumn{1}{c|}{test} &                         \\ \hline\hline
\multicolumn{1}{|c|}{\multirow{5}{*}{\lrgcut}}       & \multicolumn{1}{c|}{\multirow{3}{*}{random}}  & \multicolumn{1}{c|}{none}                                                                                                          & none                    & 33                      & \multicolumn{1}{c|}{32.1} & \multicolumn{1}{c|}{36.6} & 14.3                    & \multicolumn{1}{c|}{23.7} & \multicolumn{1}{c|}{21.3} & 21.8                    \\ \cline{3-11}
\multicolumn{1}{|c|}{}                              & \multicolumn{1}{c|}{}                         & \multicolumn{1}{c|}{target language only}                  & 100                     & \multirow{7}{*}{26}    & \multicolumn{1}{c|}{31.4} & \multicolumn{1}{c|}{33.4} & 12.6                    & \multicolumn{1}{c|}{23.2} & \multicolumn{1}{c|}{20.4} & 21.4                    \\ \cline{3-4} \cline{6-11}
\multicolumn{1}{|c|}{}                              & \multicolumn{1}{c|}{}                         & \multicolumn{1}{c|}{all (multilingual)}                                                                                 & 300                     &                         & \multicolumn{1}{c|}{26.8} & \multicolumn{1}{c|}{28.7} & 12.2                    & \multicolumn{1}{c|}{20.7} & \multicolumn{1}{c|}{18.2} & 19.7                    \\ \cline{2-4} \cline{6-11}
\multicolumn{1}{|c|}{}                              & \multicolumn{1}{c|}{\multirow{2}{*}{\xlsrcut}} & \multicolumn{1}{c|}{none}                                                                                                          & none                    &                         & \multicolumn{1}{c|}{25.4} & \multicolumn{1}{c|}{29.4} & 11.4                    & \multicolumn{1}{c|}{20.7} & \multicolumn{1}{c|}{17.9} & \textbf{19.2} \\ \cline{3-4} \cline{6-11}
\multicolumn{1}{|c|}{}                              & \multicolumn{1}{c|}{}                         & \multicolumn{1}{c|}{\multirow{2}{*}{target language only}} & 25                      &                         & \multicolumn{1}{c|}{27.6} & \multicolumn{1}{c|}{29.5} & 11.7                    & \multicolumn{1}{c|}{20.8} & \multicolumn{1}{c|}{17.7} & 19.7                    \\ \cline{1-2} \cline{4-4} \cline{6-11}
\multicolumn{1}{|c|}{\lrg}                           & \multicolumn{1}{c|}{\xlsr}                     & \multicolumn{1}{c|}{}                                                                                                              & 100                     &                         & \multicolumn{1}{c|}{26.2} & \multicolumn{1}{c|}{29.0} & 11.4                    & \multicolumn{1}{c|}{20.4} & \multicolumn{1}{c|}{\textbf{17.5}} & 19.6                    \\ \cline{1-4} \cline{6-11}
\multicolumn{1}{|c|}{\lrgcut}                        & \multicolumn{1}{c|}{\xlsrcut}                & \multicolumn{1}{c|}{\multirow{2}{*}{all (multilingual)}}                                                                & 50     &                         & \multicolumn{1}{c|}{\textbf{23.9}} & \multicolumn{1}{c|}{\textbf{27.4}} & \textbf{11.3} & \multicolumn{1}{c|}{20.6} & \multicolumn{1}{c|}{17.8} & 19.6                    \\ \cline{1-2}\cline{4-4} \cline{6-11}
\multicolumn{1}{|c|}{\lrg}                           & \multicolumn{1}{c|}{\xlsr}                     & \multicolumn{1}{c|}{}                                                                                                              &             175            &                         & \multicolumn{1}{c|}{28.1}    & \multicolumn{1}{c|}{27.8}    & 11.5                       & \multicolumn{1}{c|}{\textbf{20.1}}    & \multicolumn{1}{c|}{17.8}    & 19.5
\\ \hline
\end{tabular}
\label{table:pretrained_joined}
\vspace{-10pt}
\end{table*}

We investigate different data conditions for pre-training and present the results in \tab\ref{table:pretrained_joined}.
All these models are fine-tuned using \lrgcut and are trained until full convergence on the in-house data of the respective language.
The number of epochs for pre-training is selected based on best downstream \gls{WER} on Vietnamese.
As a baseline, we show results when fine-tuning from scratch (random initialization) in the first row.

Note that the \wvtwo model was adapted to operate on \SI{8}{\kilo\hertz} sampled data by halving the stride of the feature extractor's last convolutional layer.
\sect\ref{sec:bandwidth} studies this in more detail.
Furthermore, preliminary experiments showed that it is generally helpful to keep more encoder blocks in \lrgcutn.
However, we found 8 blocks to be a good trade-off between having a sufficiently large batch size and a model size that still fits into memory.

\textbf{Single stage pre-training:}
For both languages, pre-training on the monolingual in-house data shows relative \glspl{WERR} of mostly 2-4\%, even though no additional data is included for pre-training here.
Next, we look at models pre-trained on multilingual data.
Combining the Arabic, German and Vietnamese in-house data to do a custom multilingual pre-training clearly outperforms the monolingual baselines relative in \gls{WER} by about 14\% for Vietnamese and 11\% for German.
Simply using the \xlsrcut checkpoint directly for fine-tuning also shows gains in a straightforward way.
Moreover, it is worth noting that other ways to deal with the different bandwidth show additional gains as discussed in \sect\ref{sec:bandwidth}.

\textbf{\xlsr as pre-training initialization:}
Alternatively, we can exploit \xlsrcut by using it as an initialization for custom pre-training.
This helps both in the monolingual case with \glspl{WERR} of 10-13\% as well as the multilingual case.
In the latter, the gains are much smaller, only on Vietnamese dev the benefit is similar.
This indicates that the diverse and multilingual data used in \xlsr helps more for our pre-training on monolingual and thus less diverse data.
Additionally, we pre-train a \lrg model initialized with \xlsr and subsequently reduce it to the smaller size for fine-tuning.
This usually outperforms pre-training with the smaller \lrgcut by small margins of up to 2\% at the expense of higher resource consumption in pre-training, except for degradations on Vietnamese with \datainhouseml.
We also observe that the gains on the in-house test set are usually smaller, indicating that pre-training especially helps to obtain systems that are more robust to domain changes.

We can thus conclude that pre-training helps to improve over fine-tuning from scratch in all experiments and multilingual data for pre-training is better than (more limited) monolingual data.
Initializing the pre-training with \xlsr helps, especially in the monolingual case.
Additionally, we show that continuing the pre-training of \xlsr on custom data is a simple way to improve the results further.
Finally, the trends are similar for Vietnamese and German and the following experiments will therefore be done on Vietnamese only.

\subsection{Sampling Rate Mismatch}
\label{sec:bandwidth}

\begin{table}[!htb]
\setlength{\tabcolsep}{0.4em}
\centering
\caption{
WERs {[}\%{]} on Vietnamese for up-sampling as well as adjustments to \SI{8}{\kilo\hertz} sampling rate at different feature extractor layers.
}
\begin{tabular}{|c|c|c|ccc|}
\hline
\multirow{3}{*}{\begin{tabular}[c]{@{}c@{}}Data\\ up-sampling\\ to 16kHz\end{tabular}} & \multirow{3}{*}{\begin{tabular}[c]{@{}c@{}}NN\\ layer\end{tabular}} & \multirow{3}{*}{\begin{tabular}[c]{@{}c@{}}Pre-training \\ init/data\end{tabular}}    & \multicolumn{3}{c|}{WER {[}\%{]}}                                                          \\ \cline{4-6}
                                                                                       &                                                                     &                                                                                       & \multicolumn{2}{c|}{\hykist}                                   & \multirow{2}{*}{In-house} \\ \cline{4-5}
                                                                                       &                                                                     &                                                                                       & \multicolumn{1}{c|}{dev}           & \multicolumn{1}{c|}{test} &                           \\ \hline\hline
TensorFlow                                                                             & \multirow{2}{*}{none}                                               & \multirow{4}{*}{\begin{tabular}[c]{@{}c@{}}\xlsrcut/\\none\end{tabular}}              & \multicolumn{1}{c|}{25.6}          & \multicolumn{1}{c|}{29.4} & \textbf{11.2} \\ \cline{1-1} \cline{4-6}
FFmpeg                                                                                 &                                                                     &                                                                                       & \multicolumn{1}{c|}{24.7}          & \multicolumn{1}{c|}{28.8} & \textbf{11.2} \\ \cline{1-2} \cline{4-6}
\multirow{6}{*}{none}                                                                  & first                                                               &                                                                                       & \multicolumn{1}{c|}{25.4}          & \multicolumn{1}{c|}{29.4} & 11.4                      \\ \cline{2-2} \cline{4-6}
                                                                                       & last                                                                &                                                                                       & \multicolumn{1}{c|}{27.6}          & \multicolumn{1}{c|}{31.9} & 12.4                      \\ \cline{2-3} \cline{4-6}
                                                                                       & first                                                               & \multirow{2}{*}{\begin{tabular}[c]{@{}c@{}}scratch/\\all in-house\end{tabular}}       & \multicolumn{1}{c|}{26.9}          & \multicolumn{1}{c|}{28.5} & 12.0                      \\ \cline{2-2} \cline{4-6}
                                                                                       & last                                                                &                                                                                       & \multicolumn{1}{c|}{26.8}          & \multicolumn{1}{c|}{28.7} & 12.2                      \\ \cline{2-3} \cline{4-6}
                                                                                       & first                                                               & \multirow{2}{*}{\begin{tabular}[c]{@{}c@{}c@{}}\xlsrcut /\\all in-house\end{tabular}} & \multicolumn{1}{c|}{26.9}          & \multicolumn{1}{c|}{\textbf{27.2}} & 11.8                      \\ \cline{2-2} \cline{4-6}
                                                                                       & last                                                                &                                                                                       & \multicolumn{1}{c|}{\textbf{23.9}}          & \multicolumn{1}{c|}{27.4} & 11.3                      \\ \hline
\end{tabular}
\label{table:8khz_joined}
\end{table}

One obvious challenge for the data at hand is the telephone speech, which does not match the \SI{16}{\kilo\hertz} sampling rate used for the original model \cite{facebook2020wav2vec2}.
The feature extractor can be adapted in order to operate on \SI{8}{\kilo\hertz} sampled data while still outputting representations with the same frame shift of \SI{20}{\milli\second} by halving the stride of a convolutional layer from the feature extractor.
If the stride is even, the adjustment is straight-forward.
However, for the stride of 5 in the first layer, we alternate between moving the kernel by 2 and 3 frames, effectively halving the stride.
This can be implemented by nearest neighbor up-sampling followed by a convolution with stride 5 and dilation 2, losing the advantage of lower complexity for cases without up-sampling though.
We also halve the kernel size by summing pairs of adjacent weights here.

Another trivial solution to counter this is re-sampling all data to \SI{16}{\kilo\hertz} as done e.g. in \cite{robust_wav2vec2}.
We use both up-sampling as part of the model in TensorFlow as well as creating an up-sampled copy of the data using FFmpeg for experiments where no \SI{8}{\kilo\hertz} sampled data was seen in pre-training.

The results are shown in \tab\ref{table:8khz_joined}.
When directly fine-tuning \xlsrcut, up-sampling works better than adjusting the feature extractor, likely because no \SI{8}{\kilo\hertz} sampled data was seen in pre-training.
When modifying strides, it is best to do it in the first layer keeping the same receptive field for all layers.
When pre-training on \SI{8}{\kilo\hertz} sampled data from scratch, the results are very close for all modified layers.
However, it is better to modify the last layer when doing continued pre-training with \xlsrcut.
Results with intermediate layers are between first and last layers and not shown here.

\subsection{Freezing Schedule and Partial Freezing}
During fine-tuning in \cite{facebook2020wav2vec2}, initially only the output layer is updated while the feature extractor is not trained at all.
The experiments so far always updated all weights based on the intuition that modifying the strides requires further training also in the feature extractor.
Further experimentation not presented in detail here showed, that this helps for direct fine-tuning of \xlsrcut, however, keeping the freezing schedule is better when using checkpoints pre-trained on \SI{8}{\kilo\hertz} sampled data.

The choice of using encoder blocks 1-8 above was motivated by memory limitations in fine-tuning.
Alternatively, we propose partial freezing where we use more blocks but only train the last 8 and freeze the rest of the model including the feature extractor.
The results are depicted in \tab\ref{table:partial_freezing}.
The first row with 8 blocks is the baseline where all blocks are trainable.
We can observe that using more blocks indeed helps to improve the performance at the expense of a larger model but with little increase in training time and memory consumption during fine-tuning.
Notably, the best results are obtained with 18 blocks, not with the full model.

\begin{table}[!t]
\setlength{\tabcolsep}{0.4em}
\centering
\caption{
WERs {[}\%{]} on Vietnamese after continued pre-training with \lrg on \datainhouseml. Fine-tuning is done on \datainhouse[V] using different cut-outs with partial freezing, optimizing only the last eight Transformer blocks.
}
\begin{tabular}{|c|ccc|}
\hline
\multirow{3}{*}{\begin{tabular}[c]{@{}c@{}}Fine-tuning\\ cut-out\\\lrgcutn\end{tabular}} & \multicolumn{3}{c|}{WER {[}\%{]}}                                                 \\ \cline{2-4}
                                                                                         & \multicolumn{2}{c|}{\hykist}                          & \multirow{2}{*}{In-house} \\ \cline{2-3}
                                                                                         & \multicolumn{1}{c|}{dev}  & \multicolumn{1}{c|}{test} &                           \\ \hline\hline
8                                                                                        & \multicolumn{1}{c|}{28.5} & \multicolumn{1}{c|}{27.6} & 11.7                      \\ \hline
11                                                                                       & \multicolumn{1}{c|}{21.8} & \multicolumn{1}{c|}{25.2} & 10.4                      \\ \hline
13                                                                                       & \multicolumn{1}{c|}{20.0} & \multicolumn{1}{c|}{23.1} & 10.1                      \\ \hline
15                                                                                       & \multicolumn{1}{c|}{\textbf{18.6}} & \multicolumn{1}{c|}{20.2} & 10.1                      \\ \hline
18                                                                                       & \multicolumn{1}{c|}{\textbf{18.6}} & \multicolumn{1}{c|}{\textbf{19.4}} & \textbf{9.6} \\ \hline
21                                                                                       & \multicolumn{1}{c|}{19.3} & \multicolumn{1}{c|}{20.5} & 9.8                       \\ \hline
24                                                                                       & \multicolumn{1}{c|}{21.4} & \multicolumn{1}{c|}{22.3} & 10.7                      \\ \hline
\end{tabular}
\label{table:partial_freezing}
\end{table}

\subsection{Exploitation of Labeled Target Domain Data}
So far, only out-of-domain data was used for training.
As described in \sect\ref{sec:data}, a small dataset for adaptation to the target medical domain \datahykist[V] exists.
In \tab\ref{table:hyk_pipeline}, we investigate how to exploit this data best.
First, we observe that it is possible to train a model with only labeled \hykist data achieving \glspl{WER} well below 40\% after continued pre-training on \datainhouse[VGA] which is not the case when directly fine-tuning \xlsrcut.
Regarding the fine-tuning schedule, it is best to fine-tune on \datainhouse[V] first and then adapt on \datahykist[V].
Using \xlsrcut as initialization in custom pre-training helps for all fine-tuning data conditions.
In contrast, an additional pre-training stage on \datahykist[V] does not change the results significantly.

\begin{table}[!t]
\setlength{\tabcolsep}{0.4em}
\centering
\caption{
  WERs {[}\%{]} on Vietnamese for different applications of the \hykist train data \datahykist[V] (\SI{0.8}{\hour}) in both pre- and	fine-tuning.
  Pre-training was done using the \lrgcut architecture, either just using \xlsrcut init, continuing the pre-training on \datainhouse[VGA] or further continuing the pre-training on \datahykist[V].
  Fine-tuning was done either on \datainhouse[V] or \datahykist[V] or on both, either jointly ("+") or subsequently in two stages ("$\to$").
}
\begin{tabular}{|c|c|c|cc|}
\hline
\multirow{3}{*}{\begin{tabular}[c]{@{}c@{}}NN\\layer\end{tabular}}  & \multirow{3}{*}{\begin{tabular}[c]{@{}c@{}}Data for\\continued\\pre-training\end{tabular}} & \multirow{3}{*}{\begin{tabular}[c]{@{}c@{}}Fine-tuning data\\(target language\\only)\end{tabular}} & \multicolumn{2}{c|}{WER {[}\%{]}} \\ \cline{4-5} 
                                                                    &                                                                             &                                       & \multicolumn{2}{c|}{\hykist}                          \\ \cline{4-5}
                                                                    &                                                                             &                                       & \multicolumn{1}{c|}{dev}  & \multicolumn{1}{c|}{test} \\ \hline\hline
\multirow{4}{*}{first}                                              & \multirow{4}{*}{none}                                                       & in-house                              & \multicolumn{1}{c|}{25.4} & \multicolumn{1}{c|}{29.4} \\ \cline{3-5}
                                                                    &                                                                             & \hykist                               & \multicolumn{1}{c|}{80.0} & \multicolumn{1}{c|}{82.6} \\ \cline{3-5}
                                                                    &                                                                             & in-h. + \hykist                       & \multicolumn{1}{c|}{21.1} & \multicolumn{1}{c|}{26.6} \\ \cline{3-5}
                                                                    &                                                                             & in-h.$\to$\hykist                     & \multicolumn{1}{c|}{21.3} & \multicolumn{1}{c|}{26.1} \\ \hline
\multirow{5}{*}{last}                                               & \multirow{4}{*}{\begin{tabular}[c]{@{}c@{}}all\\in-house\end{tabular}}      & in-house                              & \multicolumn{1}{c|}{23.9} & \multicolumn{1}{c|}{27.4} \\ \cline{3-5}
                                                                    &                                                                             & \hykist                               & \multicolumn{1}{c|}{35.4} & \multicolumn{1}{c|}{37.5} \\ \cline{3-5}
                                                                    &                                                                             & in-h. + \hykist                       & \multicolumn{1}{c|}{20.2} & \multicolumn{1}{c|}{24.6} \\ \cline{3-5}
                                                                    &                                                                             & \multirow{2}{*}{in-h.$\to$\hykist}    & \multicolumn{1}{c|}{\textbf{19.6}} & \multicolumn{1}{c|}{23.4} \\ \cline{2-2} \cline{4-5}
                                                                    & all in-h.$\to$\hykist                                                       &                                       & \multicolumn{1}{c|}{\textbf{19.6}} & \multicolumn{1}{c|}{\textbf{23.2}} \\ \hline
\end{tabular}
\label{table:hyk_pipeline}
\end{table}

Finally, we combine the above insights by using the continued pre-training with the \lrg model adapted in the last stride, add the freezing schedule for fine-tuning and pick the best cut-out from \tab\ref{table:partial_freezing}.
Moreover, we study whether only updating the last $M$ blocks during supervised \hykist adaptation can be beneficial and present the results in \tab\ref{table:best_pipline_partial_freezing}.
While this was more beneficial for smaller models in preliminary experiments, it only yields marginal gains on test for \lrgcuteighteen.

\begin{table}[!t]
\setlength{\tabcolsep}{0.4em}
\centering
\vspace{-4mm}
\caption{
WERs {[}\%{]} on Vietnamese after adapting the best system from \tab\ref{table:partial_freezing} (\lrgcuteighteen) on \hykist (\datahykist[V]) in fine-tuning, where only the last $M$ Transformer blocks are trainable.
}
\begin{tabular}{|c|cc|}
\hline
\multirow{3}{*}{\begin{tabular}[c]{@{}c@{}}Number of\\ trainable Trans-\\ former blocks\end{tabular}} & \multicolumn{2}{c|}{WER {[}\%{]}} \\ \cline{2-3}
  & \multicolumn{2}{c|}{\hykist}     \\ \cline{2-3}
  & \multicolumn{1}{c|}{dev}  & test \\ \hline\hline
8 & \multicolumn{1}{c|}{\textbf{15.9}} & 18.3 \\ \hline
6 & \multicolumn{1}{c|}{16.1} & \textbf{18.2} \\ \hline
4 & \multicolumn{1}{c|}{16.3} & \textbf{18.2} \\ \hline
2 & \multicolumn{1}{c|}{16.8} & 18.5 \\ \hline
\end{tabular}
\label{table:best_pipline_partial_freezing}
\end{table}

\section{Conclusion}
In this work, we investigate how to use large pre-trained models efficiently for \gls{ASR} on a challenging low resource task.
The sampling rate mismatch is addressed by modifications of the feature extractor.
We reduce the model size effectively by using intermediate representations of the pre-trained model in fine-tuning.
The proposed freezing scheme allows rel. \glspl{WERR} of at least 18\% without requiring more \gls{GPU} memory in training.
Moreover, we experiment with different data conditions showing that the multi-stage approach outperforms the vanilla application of \xlsr by a 5\% rel. \gls{WERR} and additional adaption on \SI{0.8}{\hour} of in-domain data is still beneficial for pre-trained models.
The final model outperforms the supervised baseline without adaptation \cite{rwth2022hykist_baseline} by 48\% rel. in \gls{WER}.

\section{Acknowledgements}
\label{sec:acknowledgement}
This work was partially supported by the project HYKIST funded by the German Federal Ministry of Health on the basis of a decision of the German Federal Parliament (Bundestag) under funding ID ZMVI1-2520DAT04A.

\vfill\pagebreak

\bibliographystyle{IEEEbib-abbrev}
\bibliography{mybib}

\begin{thebibliography}{10}

\bibitem{cui2015babel}
J. Cui, B. Kingsbury, B. Ramabhadran, A. Sethy, K. Audhkhasi, X. Cui, E.
  Kislal, L. Mangu, M. Nussbaum-Thom, M. Picheny, et~al.,
\newblock ``Multilingual representations for low resource speech recognition
  and keyword search,''
\newblock in {\em Proc. ASRU}. IEEE, 2015, pp. 259--266.

\bibitem{mohamed2022representation_review}
A. Mohamed, H.-y. Lee, L. Borgholt, J.~D. Havtorn, J. Edin, C. Igel, K.
  Kirchhoff, S.-W. Li, K. Livescu, L. Maal{\o}e, et~al.,
\newblock ``Self-supervised speech representation learning: A review,''
\newblock {\em IEEE Journal of Selected Topics in Signal Processing}, 2022.

\bibitem{deepmind2018cpc}
A.~v.~d. Oord, Y. Li, and O. Vinyals,
\newblock ``Representation learning with contrastive predictive coding,''
\newblock {\em arXiv preprint arXiv:1807.03748}, 2018.

\bibitem{facebook2019wav2vec}
S. Schneider, A. Baevski, R. Collobert, and M. Auli,
\newblock ``wav2vec: Unsupervised pre-training for speech recognition,''
\newblock in {\em Proc. Interspeech}, Graz, Austria, Sept. 2019, pp.
  3465--3469.

\bibitem{facebook2020wav2vec2}
A. Baevski, Y. Zhou, A. Mohamed, and M. Auli,
\newblock ``wav2vec 2.0: A framework for self-supervised learning of speech
  representations,''
\newblock in {\em Advances in neural information processing systems}, 2020,
  vol.~33, pp. 12449--12460.

\bibitem{facebook2021hubert}
W.-N. Hsu, Y.-H.~H. Tsai, B. Bolte, R. Salakhutdinov, and A. Mohamed,
\newblock ``{HuBERT}: How much can a bad teacher benefit {ASR} pre-training?,''
\newblock in {\em Proc. ICASSP}. IEEE, 2021, pp. 6533--6537.

\bibitem{facebook2022wav2vecaug}
A. Sriram, M. Auli, and A. Baevski,
\newblock ``Wav2vec-aug: Improved self-supervised training with limited data,''
\newblock in {\em Proc. Interspeech}, 2022, pp. 4950--4954.

\bibitem{livescu2021wav2vec_analysis}
A. Pasad, J.-C. Chou, and K. Livescu,
\newblock ``Layer-wise analysis of a self-supervised speech representation
  model,''
\newblock in {\em Proc. ASRU}. IEEE, 2021, pp. 914--921.

\bibitem{robust_wav2vec2}
W.-N. Hsu, A. Sriram, A. Baevski, T. Likhomanenko, Q. Xu, V. Pratap, J. Kahn,
  A. Lee, R. Collobert, G. Synnaeve, and M. Auli,
\newblock ``Robust wav2vec 2.0: Analyzing domain shift in self-supervised
  pre-training,''
\newblock 2021, pp. 721--725.

\bibitem{facebook2020xlsr}
A. Conneau, A. Baevski, R. Collobert, A. Mohamed, and M. Auli,
\newblock ``Unsupervised cross-lingual representation learning for speech
  recognition,''
\newblock in {\em Proc. Interspeech}, 2020, pp. 2426--2430.

\bibitem{microsoft2021unispeech}
C. Wang, Y. Wu, Y. Qian, K. Kumatani, S. Liu, F. Wei, M. Zeng, and X. Huang,
\newblock ``Unispeech: Unified speech representation learning with labeled and
  unlabeled data,''
\newblock in {\em International Conference on Machine Learning}. PMLR, 2021,
  pp. 10937--10947.

\bibitem{zhang2021xlst}
Z.-Q. Zhang, Y. Song, M.-H. Wu, X. Fang, and L.-R. Dai,
\newblock ``{XLST}: Cross-lingual self-training to learn multilingual
  representation for low resource speech recognition,''
\newblock {\em arXiv preprint arXiv:2103.08207}, 2021.

\bibitem{google2022just}
J. Bai, B. Li, Y. Zhang, A. Bapna, N. Siddhartha, K.~C. Sim, and T.~N. Sainath,
\newblock ``Joint unsupervised and supervised training for multilingual
  {ASR},''
\newblock in {\em Proc. ICASSP}. IEEE, 2022, pp. 6402--6406.

\bibitem{tuske2014multilingual}
Z. T{\"u}ske, P. Golik, D. Nolden, R. Schl{\"u}ter, and H. Ney,
\newblock ``Data augmentation, feature combination, and multilingual neural
  networks to improve {ASR} and {KWS} performance for low-resource languages,''
\newblock in {\em Proc. Interspeech}, Singapore, Sept. 2014, pp. 1420--1424.

\bibitem{2021_shrinking_bigfoot}
Z. Peng, A. Budhkar, I. Tuil, J. Levy, P. Sobhani, R. Cohen, and J. Nassour,
\newblock ``Shrinking bigfoot: Reducing wav2vec 2.0 footprint,''
\newblock in {\em Proceedings of the Second Workshop on Simple and Efficient
  Natural Language Processing}, Virtual, Nov. 2021, pp. 134--141, Association
  for Computational Linguistics.

\bibitem{taiwan2022distilhubert}
H.-J. Chang, S.-w. Yang, and H.-y. Lee,
\newblock ``Distil{HuBERT:} speech representation learning by layer-wise
  distillation of hidden-unit {BERT},''
\newblock in {\em Proc. ICASSP}. IEEE, 2022, pp. 7087--7091.

\bibitem{rwth2022hykist_baseline}
C. L\"uscher, M. Zeineldeen, Z. Yang, P. Vieting, K. Le-Duc, W. Wang, R.
  Schl\"uter, and H. Ney,
\newblock ``Development of hybrid {ASR} systems for low resource medical domain
  conversational telephone speech,''
\newblock {\em arXiv preprint arXiv:2210.13397}, 2022.

\bibitem{vieting2021waveform}
P. Vieting, C. L{\"u}scher, W. Michel, R. Schl{\"u}ter, and H. Ney,
\newblock ``On architectures and training for raw waveform feature extraction
  in {ASR},''
\newblock in {\em Proc. ASRU}. IEEE, 2021, pp. 267--274.

\bibitem{doetsch2016returnn}
P. Doetsch, A. Zeyer, P. Voigtlaender, I. Kulikov, R. Schl{\"u}ter, and H. Ney,
\newblock ``{RETURNN}: the {RWTH} extensible training framework for universal
  recurrent neural networks,''
\newblock in {\em Proc. ICASSP}, New Orleans, {LA}, {USA}, 2017.

\bibitem{facebook2019fairseq}
M. Ott, S. Edunov, A. Baevski, A. Fan, S. Gross, N. Ng, D. Grangier, and M.
  Auli,
\newblock ``fairseq: A fast, extensible toolkit for sequence modeling,''
\newblock in {\em Proceedings of the 2019 Conference of the North American
  Chapter of the Association for Computational Linguistics (Demonstrations)},
  2019, pp. 48--53.

\bibitem{rybach2011rasr}
D. Rybach, S. Hahn, P. Lehnen, D. Nolden, M. Sundermeyer, Z. T{\"u}ske, S.
  Wiesler, R. Schl{\"u}ter, and H. Ney,
\newblock ``{RASR} - the {RWTH Aachen University} open source speech
  recognition toolkit,''
\newblock in {\em Proc. ASRU}, Waikoloa, HI, USA, Dec. 2011.

\end{thebibliography}

\end{document}